# Self-Dual Codes over $\mathbb{Z}_2 \times \mathbb{Z}_4$

J. Borges, S.T. Dougherty and C. Fernández-Córdoba

**Abstract**

Self-dual codes over $\mathbb{Z}_2 \times \mathbb{Z}_4$ are subgroups of $\mathbb{Z}_2^\alpha \times \mathbb{Z}_4^\beta$ that are equal to their orthogonal under an inner-product that relates to the binary Hamming scheme. Three types of self-dual codes are defined. For each type, the possible values $\alpha, \beta$ such that there exist a code $\mathcal{C} \subseteq \mathbb{Z}_2^\alpha \times \mathbb{Z}_4^\beta$ are established. Moreover, the construction of a $\mathbb{Z}_2\mathbb{Z}_4$-linear code for each type and possible pair $(\alpha, \beta)$ is given. Finally, the standard techniques of invariant theory are applied to describe the weight enumerators for each type.

**Index Terms**

Self-dual codes, Type I codes, Type II codes and $\mathbb{Z}_2\mathbb{Z}_4$-additive codes.Self-dual codes, Type I codes, Type II codes and $\mathbb{Z}_2\mathbb{Z}_4$-additive codes.

## I. INTRODUCTION

Let $\mathbb{Z}_2$ and $\mathbb{Z}_4$ be the ring of integers modulo 2 and modulo 4, respectively. Let $\mathbb{Z}_2^n$ be the set of all binary vectors of length $n$ and let $\mathbb{Z}_4^n$ be the set of all quaternary $n$-tuples. Any non-empty subset $C$ of $\mathbb{Z}_2^n$ is a binary code and a subgroup of $\mathbb{Z}_2^n$ is called a *binary linear code* or a $\mathbb{Z}_2$-*linear code*. Equivalently, any non-empty subset $\mathcal{C}$ of $\mathbb{Z}_4^n$ is a quaternary code and a subgroup of $\mathbb{Z}_4^n$ is called a *quaternary linear code*.

According to the definition given by Delsarte in 1973 (see [10]), additive codes are subgroups of the underlying abelian group in a translation association scheme. In the special case of a binary Hamming scheme, that is, when the underlying abelian group is of order $2^n$, the only structures for the abelian group are those of the form $\mathbb{Z}_2^\alpha \times \mathbb{Z}_4^\beta$, with $\alpha + 2\beta = n$ ([9]). Therefore,

This work has been partially supported by the Spanish MEC Grants MTM2009-08435 and PCI2006-A7-0616 and Catalan DURSI Grant 2009 SGR 1224.

S.T. Dougherty is at Department of Mathematics, University of Scranton, Scranton, PA 18510, USA (email: doughertys1@scranton.edu

J. Borges and C. Fernández-Córdoba are members of the Department of Information and Communications Engineering, Universitat Autònoma de Barcelona, 08193-Bellaterra, Spain (email: {jborges, cfernandez}@deic.uab.cat).

the subgroups $\mathcal{C}$ of $\mathbb{Z}_2^\alpha \times \mathbb{Z}_4^\beta$ are the only additive codes in a binary Hamming scheme. In order to distinguish them from additive codes over finite fields (see [1], [2], [3], [13]), we will hereafter call them $\mathbb{Z}_2\mathbb{Z}_4$-*additive codes* (see [4], [5], [6], [15], [16], [12]). If $\mathcal{C} \in \mathbb{Z}_2^\alpha \times \mathbb{Z}_4^\beta$ is a $\mathbb{Z}_2\mathbb{Z}_4$-additive code and $\mathbf{v} \in \mathcal{C}$, then $\mathbf{v} = (x, y)$ where $x = (x_1, \ldots, x_\alpha) \in \mathbb{Z}_2^\alpha$ and $y = (y_1, \ldots, y_\beta) \in \mathbb{Z}_4^\beta$.

Let $\mathcal{C}$ be a $\mathbb{Z}_2\mathbb{Z}_4$-additive code, which is a subgroup of $\mathbb{Z}_2^\alpha \times \mathbb{Z}_4^\beta$. We will take an extension of the usual Gray map: $\Phi : \mathbb{Z}_2^\alpha \times \mathbb{Z}_4^\beta \longrightarrow \mathbb{Z}_2^n$, where $n = \alpha + 2\beta$, given by $\Phi(x, y) = (x, \phi(y_1), \ldots, \phi(y_\beta))$, where $\phi : \mathbb{Z}_4 \longrightarrow \mathbb{Z}_2^2$ is the usual Gray map, that is, $\phi(0) = (0, 0)$, $\phi(1) = (0, 1)$, $\phi(2) = (1, 1)$, $\phi(3) = (1, 0)$. This Gray map is an isometry which transforms Lee distances defined in $\mathbb{Z}_2^\alpha \times \mathbb{Z}_4^\beta$ to Hamming distances defined in $\mathbb{Z}_2^{\alpha+2\beta}$.

Let $v_1 \in \mathbb{Z}_2^n$ and $v_2 \in \mathbb{Z}_4^\beta$. Denote by $wt_H(v_1)$ the Hamming weight of $v_1$ and $wt_L(v_2)$ the Lee weight of $v_2$. For a vector $\mathbf{v} = (v_1, v_2) \in \mathbb{Z}_2^\alpha \times \mathbb{Z}_4^\beta$, define the weight of $\mathbf{v}$, denoted by $wt(\mathbf{v})$, as $wt_H(v_1) + wt_L(v_2)$, or equivalently, the Hamming weight of $\Phi(\mathbf{v})$.

Since $\mathcal{C}$ is a subgroup of $\mathbb{Z}_2^\alpha \times \mathbb{Z}_4^\beta$, it is also isomorphic to an abelian structure $\mathbb{Z}_2^\gamma \times \mathbb{Z}_4^\delta$. Therefore, $\mathcal{C}$ is of type $2^\gamma 4^\delta$ as a group, it has $|\mathcal{C}| = 2^{\gamma+2\delta}$ codewords and the number of order two codewords in $\mathcal{C}$ is $2^{\gamma+\delta}$. Let $X$ (respectively $Y$) be the set of $\mathbb{Z}_2$ (respectively $\mathbb{Z}_4$) coordinate positions, so $|X| = \alpha$ and $|Y| = \beta$. Unless otherwise stated, the set $X$ corresponds to the first $\alpha$ coordinates and $Y$ corresponds to the last $\beta$ coordinates. Call $\mathcal{C}_X$ (respectively $\mathcal{C}_Y$) the punctured code of $\mathcal{C}$ by deleting the coordinates outside $X$ (respectively $Y$). Let $\mathcal{C}_b$ be the subcode of $\mathcal{C}$ which contains all order two codewords and let $\kappa$ be the dimension of $(\mathcal{C}_b)_X$, which is a binary linear code. For the case $\alpha = 0$, we will write $\kappa = 0$. Considering all these parameters, we will say that $\mathcal{C}$ (or equivalently $C = \Phi(\mathcal{C})$) is of type $(\alpha, \beta; \gamma, \delta; \kappa)$.

*Definition 1:* Let $\mathcal{C}$ be a $\mathbb{Z}_2\mathbb{Z}_4$-additive code, which is a subgroup of $\mathbb{Z}_2^\alpha \times \mathbb{Z}_4^\beta$. We say that the binary image $C = \Phi(\mathcal{C})$ is a $\mathbb{Z}_2\mathbb{Z}_4$-*linear code* of binary length $n = \alpha + 2\beta$ and type $(\alpha, \beta; \gamma, \delta; \kappa)$, where $\gamma$, $\delta$ and $\kappa$ are defined as above.

Note that $\mathbb{Z}_2\mathbb{Z}_4$-additive codes are a generalization of binary linear codes and quaternary linear codes. When $\beta = 0$, the binary code $C = \mathcal{C}$ corresponds to a binary linear code. On the other hand, when $\alpha = 0$, the $\mathbb{Z}_2\mathbb{Z}_4$-additive code $\mathcal{C}$ is a quaternary linear code.

Let $\mathcal{C}$ be a $\mathbb{Z}_2\mathbb{Z}_4$-additive code. Although a $\mathbb{Z}_2\mathbb{Z}_4$-additive code $\mathcal{C}$ is not a free module, every codeword is uniquely expressible in the form

$$\sum_{i=1}^{\gamma} \lambda_i \mathbf{u}_i + \sum_{j=1}^{\delta} \mu_j \mathbf{v}_j,$$

where $\lambda_i \in \mathbb{Z}_2$ for $1 \leq i \leq \gamma$, $\mu_j \in \mathbb{Z}_4$ for $1 \leq j \leq \delta$ and $\mathbf{u}_i, \mathbf{v}_j$ are vectors in $\mathbb{Z}_2^\alpha \times \mathbb{Z}_4^\beta$ of order two and four, respectively. The vectors $\mathbf{u}_i, \mathbf{v}_j$ give us a generator matrix $\mathcal{G}$ of size $(\gamma + \delta) \times (\alpha + \beta)$ for the code $\mathcal{C}$. Moreover, we can write $\mathcal{G}$ as

$$\mathcal{G} = \left( \begin{array}{c|c} B_1 & 2B_3 \\ B_2 & Q \end{array} \right), \tag{1}$$

where $B_1, B_2, B_3$ are matrices over $\mathbb{Z}_2$ of size $\gamma \times \alpha$, $\delta \times \alpha$ and $\gamma \times \beta$, respectively; and $Q$ is a matrix over $\mathbb{Z}_4$ of size $\delta \times \beta$ with quaternary row vectors of order four.

Let $I_k$ be the identity matrix of size $k \times k$. The following theorem gives a canonical generator matrix for $\mathbb{Z}_2\mathbb{Z}_4$-codes.

*Theorem 1 ([6]):* Let $\mathcal{C}$ be a $\mathbb{Z}_2\mathbb{Z}_4$-additive code of type $(\alpha, \beta; \gamma, \delta; \kappa)$. Then, $\mathcal{C}$ is permutation equivalent to a $\mathbb{Z}_2\mathbb{Z}_4$-additive code with canonical generator matrix of the form

$$\mathcal{G}_S = \left( \begin{array}{cc|ccc} I_\kappa & T_b & 2T_2 & \mathbf{0} & \mathbf{0} \\ \mathbf{0} & \mathbf{0} & 2T_1 & 2I_{\gamma-\kappa} & \mathbf{0} \\ \mathbf{0} & S_b & S_q & R & I_\delta \end{array} \right), \tag{2}$$

where $T_b, S_b$ are matrices over $\mathbb{Z}_2$; $T_1, T_2, R$ are matrices over $\mathbb{Z}-4$ with all entries in $\{0, 1\} \subset \mathbb{Z}_4$; and $S_q$ is a matrix over $\mathbb{Z}_4$.

Self-duality for binary and quaternary linear codes has been extensively studied, see [17] for a complete description and an extensive bibliography. For the quaternary case, the binary Gray images are also very interesting since these codes are *formally self-dual* in the sense that their Hamming weight enumerators are invariant under MacWilliams transform. Therefore, a next logical step is to study self-dual $\mathbb{Z}_2\mathbb{Z}_4$-additive codes and their binary images.

The concept of duality for $\mathbb{Z}_2\mathbb{Z}_4$-additive codes were also studied in [6], where the appropriate inner product for any two vectors $\mathbf{u}, \mathbf{v} \in \mathbb{Z}_2^\alpha \times \mathbb{Z}_4^\beta$ is defined as

$$\langle \mathbf{u}, \mathbf{v} \rangle = 2(\sum_{i=1}^{\alpha} u_i v_i) + \sum_{j=\alpha+1}^{\alpha+\beta} u_j v_j \in \mathbb{Z}_4.$$

Then, the *additive dual code* of $\mathcal{C}$, denoted by $\mathcal{C}^\perp$, is defined in the standard way

$$\mathcal{C}^\perp = \{\mathbf{v} \in \mathbb{Z}_2^\alpha \times \mathbb{Z}_4^\beta \mid \langle \mathbf{u}, \mathbf{v} \rangle = 0 \text{ for all } \mathbf{u} \in \mathcal{C}\}.$$

$\mathcal{C}$ is called self-orthogonal if $\mathcal{C}^\perp \subseteq \mathcal{C}$ and self-dual if $\mathcal{C}^\perp = \mathcal{C}$. If $C = \phi(\mathcal{C})$, the binary code $\Phi(\mathcal{C}^\perp)$ is denoted by $C_\perp$ and called the $\mathbb{Z}_2\mathbb{Z}_4$-dual code of $C$. In [6] it was proved that the additive dual code $\mathcal{C}^\perp$, which is also a $\mathbb{Z}_2\mathbb{Z}_4$-additive code, is of type $(\alpha, \beta; \bar{\gamma}, \bar{\delta}; \bar{\kappa})$, where

$$\begin{aligned}
\bar{\gamma} &= \alpha + \gamma - 2\kappa, \\
\bar{\delta} &= \beta - \gamma - \delta + \kappa, \\
\bar{\kappa} &= \alpha - \kappa.
\end{aligned} \tag{3}$$

Moreover, we can construct the parity-check matrix of a $\mathbb{Z}_2\mathbb{Z}_4$-additive code generated by a canonical generator matrix as in (2).

*Theorem 2 ([6]):* Let $\mathcal{C}$ be a $\mathbb{Z}_2\mathbb{Z}_4$-additive code of type $(\alpha, \beta; \gamma, \delta; \kappa)$ with canonical generator matrix (2). Then, the generator matrix of $\mathcal{C}^\perp$ is

$$\mathcal{H}_\mathcal{S} = \begin{pmatrix} T_b^t & I_{\alpha-\kappa} & 0 & 0 & 2S_b^t \\ 0 & 0 & 0 & 2I_{\gamma-\kappa} & 2R^t \\ T_2^t & 0 & I_{\beta+\kappa-\gamma-\delta} & T_1^t & -(S_q + RT_1)^t \end{pmatrix}, \tag{4}$$

where $T_b, T_2$ are matrices over $\mathbb{Z}_2$; $T_1, R, S_b$ are matrices over $\mathbb{Z}_4$ with all entries in $\{0,1\} \subset \mathbb{Z}_4$; and $S_q$ is a matrix over $\mathbb{Z}_4$. Moreover, $T_2$ and $S_b$ are obtained from the matrices of (2) with the same name, whose all entries are $\{0,1\}$, after applying the modulo 2 map and the inclusion, respectively.

Let $\mathcal{C}$ be a $\mathbb{Z}_2\mathbb{Z}_4$-linear code of type $(\alpha, \beta; \gamma, \delta; \kappa)$. Define the usual Hamming weight enumerator of $\mathcal{C}$ to be

$$W_\mathcal{C}(x, y) = \sum_{\mathbf{c} \in \mathcal{C}} x^{n-wt(\mathbf{c})} y^{wt(\mathbf{c})},$$

where $n = \alpha + 2\beta$. We know from [9], [16], [5] that for the weight enumerator defined above we have

$$W_{\mathcal{C}^\perp}(x, y) = \frac{1}{|\mathcal{C}|} W_\mathcal{C}(x + y, x - y).$$

*Lemma 1 ([6]):* Let $\mathcal{C}$ be a $\mathbb{Z}_2\mathbb{Z}_4$-additive code of type $(\alpha, \beta; \gamma, \delta; \kappa)$ and $\mathcal{C}^\perp$ its additive dual code. Then, $|\mathcal{C}||\mathcal{C}^\perp| = 2^n$, where $n = \alpha + 2\beta$.

## II. Classification of $\mathbb{Z}_2\mathbb{Z}_4$-additive self-dual codes

*Lemma 2:* If $\mathcal{C}$ is an additive self-dual code, then $\mathcal{C}$ is of type $(2\kappa, \beta; \beta + \kappa - 2\delta, \delta; \kappa)$, $|\mathcal{C}| = 2^{\kappa+\beta}$ and $|\mathcal{C}_b| = 2^{\kappa+\beta-\delta}$.

*Proof:* By (3), we have that $\alpha = 2\kappa$ and $\gamma = \beta + \kappa - 2\delta$. Since $|\mathcal{C}| = 2^{\gamma+2\delta}$ and $|\mathcal{C}_b| = 2^{\gamma+\delta}$, the result holds. ∎

*Corollary 1:* If $\mathcal{C}$ is a $\mathbb{Z}_2\mathbb{Z}_4$-additive self-dual code of type $(\alpha, \beta; \gamma, \delta; \kappa)$ and length $n$, then $n$ is even and $\alpha$ is even.

*Lemma 3:* Let $\mathcal{C} \subseteq \mathbb{Z}_2^\alpha \times \mathbb{Z}_4^\beta$ be an additive self-dual code. Let $(x, y) \in \mathcal{C}$, and denote $p(y)$ the number of odd (order four) coordinates of any vector $y \in \mathbb{Z}_4^\beta$. Thus we obtain:

(i) If $wt_H(x)$ is even, then $p(y) \equiv 0 \pmod 4$.

(ii) If $wt_H(x)$ is odd, then $p(y) \equiv 2 \pmod 4$.

*Proof:* Given a codeword $\mathbf{v} = (x, y) \in \mathcal{C}$, since $\mathbf{v}$ must be self-orthogonal, we have $\langle \mathbf{v}, \mathbf{v} \rangle = 2wt_H(x) + p(y) = 0 \in \mathbb{Z}_4$ and the statement holds. ∎

Given $j \in \{0, 1, 2, 3\}$, we denote by $\mathbf{j}^r$ the tuple $(j, j, \ldots, j)$ of length $r$. Therefore, $(\mathbf{0}^\alpha, \mathbf{2}^\beta)$ is clearly a codeword in $\mathcal{C}$.

*Lemma 4:* If $\mathcal{C}$ is an additive self-dual code, then the subcode $(\mathcal{C}_b)_X$ is a binary self-dual code.

*Proof:* By Lemma 2, the code $\mathcal{C}$ is of type $(2\kappa, \beta; \beta + \kappa - 2\delta, \delta; \kappa)$. Since for any pair of codewords $(x, y), (x', y') \in \mathcal{C}_b$ we have that $y$ and $y'$ are orthogonal quaternary vectors, $(\mathcal{C}_b)_X \subseteq (\mathcal{C}_b)_X^\perp$. Moreover, since $(\mathcal{C}_b)_X$ has dimension $\kappa$ (by definition) and is of length $2\kappa$, we have that $(\mathcal{C}_b)_X$ is binary self-dual. ∎

*Lemma 5:* Let $\mathcal{C}$ be an additive self-dual code of type $(2\kappa, \beta; \beta + \kappa - 2\delta, \delta; \kappa)$. There is an integer number $r$, $0 \leq r \leq \kappa$, such that each codeword in $\mathcal{C}_Y$ appears $2^r$ times in $\mathcal{C}$ and $|\mathcal{C}_Y| \geq 2^\beta$.

*Proof:* Consider the subcode $\bar{\mathcal{C}} = \{(x, \mathbf{0}^\beta) \in \mathcal{C}\}$. Clearly, $(\bar{\mathcal{C}})_X$ is a binary linear code. Let $r = dim(\bar{\mathcal{C}})_X$. Thus, any vector in $\mathcal{C}_Y$ appears $2^r$ times in $\mathcal{C}$. Note that $(\bar{\mathcal{C}})_X$ is also a subcode of $(\mathcal{C}_b)_X$, hence $r \leq \kappa$. Also, we have that $|\mathcal{C}| = 2^{\beta+\kappa} = |\mathcal{C}_Y| \cdot 2^r$, therefore $|\mathcal{C}_Y| \geq 2^\beta$. ∎

Let $\mathcal{C}$ be a $\mathbb{Z}_2\mathbb{Z}_4$-additive code. If $\mathcal{C} = \mathcal{C}_X \times \mathcal{C}_Y$, then $\mathcal{C}$ is called *separable*. If $\mathcal{C}$ is a separable $\mathbb{Z}_2\mathbb{Z}_4$-additive code, then the generator matrix of $\mathcal{C}$ in standard form is

$$\mathcal{G}_S = \begin{pmatrix} I_\kappa & T' & \mathbf{0} & \mathbf{0} & \mathbf{0} \\ \mathbf{0} & \mathbf{0} & 2T_1 & 2I_{\gamma-\kappa} & \mathbf{0} \\ \mathbf{0} & \mathbf{0} & S & R & I_\delta \end{pmatrix}.$$

The following theorems show some properties of separable $\mathbb{Z}_2\mathbb{Z}_4$-additive self-dual codes.

*Theorem 3:* Let $\mathcal{C}$ be a $\mathbb{Z}_2\mathbb{Z}_4$-additive self-dual code of type $(2\kappa, \beta; \beta + \kappa - 2\delta, \delta; \kappa)$. The following statements are equivalent:

(i) $\mathcal{C}_X$ is binary self-orthogonal.

(ii) $\mathcal{C}_X$ is binary self-dual.

(iii) $|\mathcal{C}_X| = 2^\kappa$.

(iv) $\mathcal{C}_Y$ is a quaternary self-orthogonal code.

(v) $\mathcal{C}_Y$ is a quaternary self-dual code.

(vi) $|\mathcal{C}_Y| = 2^\beta$.

(vii) $\mathcal{C}$ is separable.

*Proof:* $(i) \Leftrightarrow (ii)$: By Lemma 4, $|\mathcal{C}_X| \geq 2^\kappa$, thus $(i)$ and $(ii)$ are equivalent statements.

$(ii) \Leftrightarrow (iii)$: Clearly, $(ii)$ implies $(iii)$ and $(iii)$ implies $\mathcal{C}_X = (\mathcal{C}_b)_X$ and $\mathcal{C}_X$ is binary self-dual, by Lemma 4.

$(ii) \Leftrightarrow (v)$: Straightforward.

$(iv) \Leftrightarrow (v)$: By Lemma 5, $|\mathcal{C}_Y| \geq 2^\beta$, thus $(iv)$ and $(v)$ are equivalent statements.

$(ii) \Leftrightarrow (vii)$: If $\mathcal{C}_X$ is binary self-dual, then $\mathcal{C}_Y$ is quaternary self-dual, $|\mathcal{C}_X| = 2^\kappa$ and $|\mathcal{C}_Y| = 2^\beta$. Since $\mathcal{C} = 2^{\kappa+\beta}$ we have that the set of codewords in $\mathcal{C}$ is $\mathcal{C}_X \times \mathcal{C}_Y$. Reciprocally, if $\mathcal{C} = \mathcal{C}_X \times \mathcal{C}_Y$, then $(x, \mathbf{0}^\beta) \in \mathcal{C}$ for any $x \in \mathcal{C}_X$ and $\mathcal{C}_X$ must be a binary self-dual code. Also, $(\mathbf{0}^\alpha, y) \in \mathcal{C}$ for any $y \in \mathcal{C}_Y$ and $\mathcal{C}_Y$ must be a quaternary self-dual code.

$(v) \Rightarrow (vi)$: Trivial.

$(vi) \Rightarrow (iii)$: By Lemma 5, each vector in $\mathcal{C}_Y$ appears $2^\kappa$ times in $\mathcal{C}$. Thus, for any vector $x_b \in (\mathcal{C}_b)_X$, the vector $(x_b, \mathbf{0}^\beta)$ is a codeword in $\mathcal{C}_b$. This means that given any codeword $(x, y) \in \mathcal{C}$, we have that $x$ and $x_b$ are orthogonal binary vectors, for all $x_b \in (\mathcal{C}_b)_X$, since $\langle (x, y), (x_b, \mathbf{0}^\beta) \rangle = 0$. Therefore, for all $x \in \mathcal{C}_X$, $x \in (\mathcal{C}_b)_X^\perp$ and $\mathcal{C}_X \subseteq (\mathcal{C}_b)_X^\perp$. By Lemma 4, $(\mathcal{C}_b)_X^\perp = (\mathcal{C}_b)_X$, which implies $\mathcal{C}_X = (\mathcal{C}_b)_X$, hence $|\mathcal{C}_X| = 2^\kappa$. ∎

*Corollary 2:* Let $\mathcal{C}$ be a $\mathbb{Z}_2\mathbb{Z}_4$-additive self-dual code. If $\mathcal{C}_X$ has all weights doubly-even, then $\mathcal{C}$ is separable.

*Proof:* Given any pair of vectors $x, y \in \mathcal{C}_X$, we have that

$$w_H(x+y) = w_H(x) + w_H(y) - 2w_H(x \cdot y),$$

where $x \cdot y$ is the componentwise product of $x$ and $y$. Since $w_H(x+y)$, $w_H(x)$ and $w_H(y)$ are 0 modulo 4, we conclude that $w_H(x \cdot y)$ is even and then $x$ and $y$ are orthogonal. Therefore $\mathcal{C}_X$ is self-orthogonal and by Theorem 3, $\mathcal{C}$ is separable. ∎

From the above theorem, if $\mathcal{C}$ is a $\mathbb{Z}_2\mathbb{Z}_4$-additive self-dual code, then $\mathcal{C}_X$ is binary self-dual if and only if $\mathcal{C}_Y$ is quaternary self-dual. Moreover, if $\mathcal{C}$ is a separable $\mathbb{Z}_2\mathbb{Z}_4$-additive code, $\mathcal{C}_X$ is binary self-dual and $\mathcal{C}_Y$ is quaternary self-dual, then $\mathcal{C}$ is also self-dual as it is proven in the following theorem.

*Theorem 4:* If $C$ is a binary self-dual code of length $\alpha$ and $\mathcal{D}$ is a quaternary self-dual code of length $\beta$ then $C \times \mathcal{D}$ is a $\mathbb{Z}_2\mathbb{Z}_4$-additive self-dual code of length $\alpha + \beta$.

*Proof:* Let $x = (x_1, \ldots, x_\alpha) \in C$, $x' = (x'_1, \ldots, x'_\alpha) \in C$, $y = (y_1, \ldots, y_\beta) \in \mathcal{D}$ and $y' = (y'_1, \ldots, y'_\beta) \in \mathcal{D}$. Since $x, x'$ are binary orthogonal vectors and $y, y'$ are quaternary orthogonal vectors, we have

$$\langle (x,y), (x',y') \rangle = 2 \sum_{i=1}^{\alpha} x_i x'_i + \sum_{j=1}^{\beta} y_j y'_j \equiv 0 \pmod{4},$$

so the code is self-orthogonal. Then $|C|^2 = 2^\alpha$ and $|\mathcal{D}|^2 = 4^\beta$ so $|C \times \mathcal{D}|^2 = |C|^2 |\mathcal{D}|^2 = 2^\alpha 4^\beta$ and the code is self-dual. △

*Definition 2:* We say that a binary code $C$ is antipodal if for any codeword $z \in C$, then $z + \mathbf{1} \in C$. If $\mathcal{C}$ is a $\mathbb{Z}_2\mathbb{Z}_4$-additive code, we say that $\mathcal{C}$ is antipodal if $\Phi(\mathcal{C})$ is antipodal.

Clearly, a $\mathbb{Z}_2\mathbb{Z}_4$-additive code $\mathcal{C} \subseteq \mathbb{Z}_2^\alpha \times \mathbb{Z}_4^\beta$ is antipodal if and only if $(\mathbf{1}^\alpha, \mathbf{2}^\beta) \in \mathcal{C}$.

*Definition 3:* If a $\mathbb{Z}_2\mathbb{Z}_4$-additive self-dual code has odd weights, then it is said to be Type 0. If it has only even weights, then the code is said to be Type I. If all the codewords have doubly-even weight then it is said to be Type II.

We remark that applying Theorem 4 to a binary self-dual code and a quaternary self-dual code gives a Type I code. Applying Theorem 4 to a binary doubly-even self-dual code and a quaternary doubly-even code gives a Type II code.

Now, we study the relation among type, separability and antipodality.

*Proposition 1:* Let $\mathcal{C} \subseteq \mathbb{Z}_2^\alpha \times \mathbb{Z}_4^\beta$ be an additive self-dual code, then $\mathcal{C}$ is antipodal if and only if $\mathcal{C}$ is of Type I or Type II.

*Proof:* Since $(\mathbf{0}^\alpha, \mathbf{2}^\beta) \in \mathcal{C}$ and $\mathcal{C}$ is antipodal if and only if $(\mathbf{1}^\alpha, \mathbf{2}^\beta) \in \mathcal{C}$, we get that $\mathcal{C}$ is antipodal if and only if $(\mathbf{1}^\alpha, \mathbf{0}^\beta) \in \mathcal{C}$. But this condition is equivalent to say that all codewords of $\mathcal{C}_X$ ($\mathcal{C}$ restricted to the binary coordinates) are of even weight. △

*Proposition 2:* Let $\mathcal{C} \subseteq \mathbb{Z}_2^\alpha \times \mathbb{Z}_4^\beta$ be an additive self-dual code. If $\mathcal{C}$ is separable, then $\mathcal{C}$ is antipodal.

*Proof:* If $\mathcal{C}$ is separable, then $\mathcal{C} = \mathcal{C}_X \times \mathcal{C}_Y$, where $\mathcal{C}_X$ is a binary linear self-dual code and $\mathcal{C}_Y$ is a quaternary linear self-dual code. Hence, $\mathcal{C}_X$ contains the all-one vector and $\mathcal{C}_Y$ contains the all-two vector. Therefore $(\mathbf{1}^\alpha, \mathbf{2}^\beta) \in \mathcal{C}$. △

*Corollary 3:* If $\mathcal{C}$ is a $\mathbb{Z}_2\mathbb{Z}_4$-additive self-dual code of Type 0, then $\mathcal{C}$ is non-separable and non-antipodal.

*Proof:* Straightforward from the previous two propositions. △

### III. EXAMPLES

In the following examples we delete customary commas between coordinates and we use '|' to separate the binary and quaternary parts of the vectors.

*Example 1 (Type 0):* Let $\mathcal{D} = \{(00 \mid 00), (00 \mid 22), (11 \mid 02), (11 \mid 20)\}$. Then, the code $\mathcal{C}_1 = \mathcal{D} \cup (\mathcal{D} + (01 \mid 11))$ is a $\mathbb{Z}_2\mathbb{Z}_4$-additive self-dual code of type $(2, 2; 1, 1; 1)$ and has generator matrix

$$\mathcal{G}_1 = \begin{pmatrix} 11 & 20 \\ 01 & 11 \end{pmatrix}.$$

The weight enumerator of this code is

$$W_\mathcal{C}(x, y) = x^6 + 4x^3y^3 + 3x^2y^4.$$

Notice that it has vectors of odd weight, hence it is a Type 0 code and by Corollary 3 it is non-separable.

*Example 2 (Type I, separable):* A $\mathbb{Z}_2\mathbb{Z}_4$-additive self-dual code with $\alpha, \beta \geq 1$ should have $\alpha \geq 2$, since $\alpha$ must be even. A $\mathbb{Z}_2\mathbb{Z}_4$-additive self-dual code with minimum length has $\alpha = 2$, $\beta = 1$ and $2^{\kappa+\beta} = 2^{1+1} = 4$ codewords. For example, the code $\mathcal{C}_2 = \{(00 \mid 0), (00 \mid 2), (11 \mid 0), (11 \mid 2)\}$ is an additive self-dual code of type $(2, 1; 2, 0; 1)$ and has generator matrix

$$\mathcal{G}_2 = \begin{pmatrix} 11 & 0 \\ 00 & 2 \end{pmatrix}.$$

Notice that for $\alpha = 2$ and $\beta = 1$, it is not possible to have odd weight codewords. Thus, the code must be of Type I and antipodal. Also, he have that the code restricted to the quaternary coordinates is $\{0, 2\}$ which is self-dual and hence, by Theorem 3, $\mathcal{C}_2$ is separable.

*Example 3 (Type I, non-separable):* Consider the following matrices:

$$\mathcal{G}_3 = \begin{pmatrix} 1111 & 0000 \\ 0101 & 2000 \\ 0101 & 0200 \\ 0101 & 0020 \\ 0011 & 1111 \end{pmatrix} ; \quad \mathcal{G}_4 = \begin{pmatrix} 1111 & 000000 \\ 0101 & 220000 \\ 0000 & 202000 \\ 0101 & 000200 \\ 0101 & 111010 \\ 0011 & 101101 \end{pmatrix}. \quad (5)$$

The codes $\mathcal{C}_3$ and $\mathcal{C}_4$ generated by $\mathcal{G}_3$ and $\mathcal{G}_4$ respectively are non-separable Type I.

*Example 4 (Type II, separable):* Let $C$ be the extended binary Hamming code of length 8 and let $\mathcal{D}$ be the quaternary linear code generated by

$$\begin{pmatrix} 2 & 2 & 0 & 0 \\ 2 & 0 & 2 & 0 \\ 1 & 1 & 1 & 1 \end{pmatrix}.$$

$|\mathcal{D}| = 2^2 4^1 = 2^4$ which is correct size to be self-dual. Clearly, $\mathcal{D}$ is quaternary self-orthogonal and hence self-dual. On the other hand, $C$ is a binary self-dual code. Since both codes have only doubly-even weights we conclude that $\mathcal{C}_5 = C \times \mathcal{D}$ is a Type II separable code with the minimum possible length.

*Example 5 (Type II, non-separable):* The code $\mathcal{C}_6$ generated by the following matrix is self-orthogonal and of type $2^6 4^1$, thus $|\mathcal{C}_6| = 2^8$ and $\mathcal{C}_6$ is self-dual.

$$\begin{pmatrix} 10010110 & 0000 \\ 01001110 & 0000 \\ 00100111 & 0000 \\ 00000110 & 2000 \\ 00000110 & 0200 \\ 00000110 & 0020 \\ 00011011 & 1111 \end{pmatrix}.$$

Clearly it is non-separable, since $(\mathcal{C}_6)_X$ is not self-orthogonal. On the other hand, it can be checked that all weights are doubly-even.

## IV. EXISTENCE CONDITIONS

The following lemma is easily proven.

*Lemma 6:* If $\mathcal{C}$ is a $\mathbb{Z}_2\mathbb{Z}_4$-additive self-dual code of type $(\alpha, \beta; \gamma, \delta; \kappa)$ and $\mathcal{D}$ is a $\mathbb{Z}_2\mathbb{Z}_4$-additive self-dual code of type $(\alpha', \beta'; \gamma', \delta'; \kappa')$ then $\mathcal{C} \times \mathcal{D}$ is a $\mathbb{Z}_2\mathbb{Z}_4$-additive self-dual code of type $(\alpha + \alpha', \beta + \beta'; \gamma + \gamma', \delta + \delta'; \kappa + \kappa')$.

Theorem 4 and Lemma 6 gives the following:

*Corollary 4:* There exist $\mathbb{Z}_2\mathbb{Z}_4$-additive self-dual codes of type $(\alpha, \beta; \gamma, \delta; \kappa)$ for all even $\alpha$ and all $\beta$.

The following corollary follows from Gleason's theorem.

*Corollary 5:* If $\mathcal{C}$ is a Type II $\mathbb{Z}_2\mathbb{Z}_4$-additive code of type $(\alpha, \beta; \gamma, \delta; \kappa)$, then $n = \alpha + 2\beta$ is a multiple of $8$.

*Lemma 7:* If $C$ is a binary linear self-dual Type I code, then it has a linear subcode $C_0$ such that $C_0$ contains all doubly-even codewords of $C$ and $|C_0| = |C|/2$.

*Proof:* Define the application $f : C \longrightarrow \mathbb{Z}_2$, such that $f(x) = 0$ if $x$ is doubly-even and $f(x) = 1$ if $x$ is single-even, for all $x \in C$. Taken into account that any pair of codewords must be orthogonal, it is easily verified that $f$ is a group morphism. Therefore, $C/Ker\ f$ is isomorphic to the image of $f$. Since $C$ contains single-even codewords, this image is $\mathbb{Z}_2$. It follows that $|C|/|Ker\ f| = 2$ and, clearly, $Ker\ f = C_0$. ∎

*Theorem 5:* If $\mathcal{C}$ is a Type II $\mathbb{Z}_2\mathbb{Z}_4$-additive code of type $(\alpha, \beta; \gamma, \delta; \kappa)$, then $\alpha \equiv 0 \pmod{8}$.

*Proof:* We shall prove that if $\mathcal{C}$ is a Type II $\mathbb{Z}_2\mathbb{Z}_4$-additive code then there exists a binary Type II code of length $\alpha$. It is well known that binary Type II codes only exists for lengths a multiple of $8$ (see also Corollary 5). If the code is separable then we already know that the binary part is a Type II code. Therefore we shall assume that $\mathcal{C}$ is not a separable code.

First, we consider the code $(\mathcal{C}_b)_X$. We know that this code is a binary self-dual code by Lemma 4. If it is Type II we are done, so assume that is Type I. Thus, by Lemma 7, $(\mathcal{C}_b)_X$ contains a codimension 1 linear subcode consisting of the doubly-even vectors. Call this subcode $((\mathcal{C}_b)_X)_0$.

For Type II codes and vectors $(v, w)$ in those codes we have:

$$wt_H(v) + wt_L(w) \equiv 0 \pmod{4}$$
$$2wt_H(v) + p(w) \equiv 0 \pmod{4},$$

where $p(w)$ is the number of order 4 coordinates in $w$ as in Lemma 3. The first is because the vectors are doubly-even and the second is because they are self-orthogonal (see also Lemma 3).

These equations imply that every vector either has doubly-even binary part with doubly-even many units in the quaternary part with evenly many 2s or that it has singly-even binary party with doubly-even many units in the quaternary part with oddly many 2s.

Consider the vector $(\mathbf{0}^\alpha, \mathbf{1}^\beta)$. This vector is not in $\mathcal{C}$ since there are vectors of the second type in $\mathcal{C}$, otherwise it would be separable by Corollary 2. Moreover $(\mathbf{0}^\alpha, \mathbf{1}^\beta) + (\mathbf{0}^\alpha, \mathbf{1}^\beta) \in \mathcal{C}$. Hence if we let $\mathcal{D} = \{\mathbf{v} \mid \mathbf{v} \in \mathcal{C}, \langle \mathbf{v}, (\mathbf{0}^\alpha, \mathbf{1}^\beta) \rangle = 0\}$, since $\mathcal{C}$ is index 2 in $\langle \mathcal{C}, (\mathbf{0}^\alpha, \mathbf{1}^\beta) \rangle$, we have that $\mathcal{D}$ is index 2 in $\mathcal{C}$. Moreover $\mathcal{D}$ consists precisely of those vectors as described above with doubly-even binary part.

Notice that the code $\mathcal{D}_X$ contains $((\mathcal{C}_b)_X)_0$. If $\mathcal{D}_X = ((\mathcal{C}_b)_X)_0$ then $\mathcal{C}_X = (\mathcal{C}_b)_X$ and the code is separable. So $((\mathcal{C}_b)_X)_0$ is strictly contained in $\mathcal{D}_X$ giving that the dimension of $\mathcal{D}_X$ is at least the dimension of $(\mathcal{C}_b)_X$. Now $\mathcal{D}_X$ is a binary linear code consisting of only doubly-even vectors. Hence, as in Corollary 2, $D_X$ is a linear, doubly-even, self-orthogonal code with dimension at least the dimension of a self-dual code, hence it must be a Type II code giving that $\alpha$ must be a multiple of 8. ■

*Theorem 6:* Let $\mathcal{C}$ be a self-dual $\mathbb{Z}_2\mathbb{Z}_4$-additive code of length $n$. Let $\mathbf{v}$ be any self-orthogonal vector that is not in $\mathcal{C}$. Let $\mathcal{C}' = \{\mathbf{w} \mid \mathbf{w} \in \mathcal{C}, \langle \mathbf{w}, \mathbf{v} \rangle = 0\}$. Then $\mathcal{D} = \langle \mathcal{C}', \mathbf{v} \rangle$ is a self-dual code of length $n$. Moreover, if $\mathbf{v}$ has odd weight then $D$ is a Type 0 code.

*Proof:* Consider the additive code $\langle \mathcal{C}, \mathbf{v} \rangle$. The index of $\mathcal{C}'$ in $\mathcal{D}$ is the same as the index of $\mathcal{C}$ in $\langle \mathcal{C}, \mathbf{v} \rangle$, hence $|\mathcal{D}||\mathcal{C}| = |\langle \mathcal{C}, \mathbf{v} \rangle||\mathcal{C}'|$. Note that $\langle \mathcal{C}, \mathbf{v} \rangle^\perp = \mathcal{C}^\perp \cap \langle \mathbf{v} \rangle^\perp = \mathcal{C} \cap \langle \mathbf{v} \rangle^\perp = \mathcal{C}'$ and therefore $|\mathcal{D}||\mathcal{C}|$ is the cardinality of the ambient space. Since $\mathcal{D}$ is self-orthogonal by contruction and it has the cardinality of a self-dual code, then it is a self-dual code of the same length as $\mathcal{C}$.

If $\mathbf{v}$ has odd weight then there is at least one odd weight vector in $\mathcal{D}$ hence is Type 0. △

The construction of the self-dual codes in Theorem 6 is known as the neighbor construction. It is a direct generalization of the technique used for codes over fields and rings. Its primary use is that it gives a simple construction technique to produce Type 0 codes.

Denote by $\langle \cdot, \cdot \rangle_2$ and $\langle \cdot, \cdot \rangle_4$ the standard inner binary and quaternary product, respectively.

*Lemma 8:* If $\mathcal{C}$ is a non-separable $\mathbb{Z}_2\mathbb{Z}_4$-additive code, then there exists two codewords $(v, w)$ and $(v', w')$ with $\langle v, v' \rangle_2 = 1$ and $\langle w, w' \rangle_4 = 2$.

July 4, 2018    DRAFT

*Proof:* If $\langle v, v'\rangle_2 = 0$ then $2\langle v, v'\rangle_2 = 0$ and so $\langle w, w'\rangle_4 = 0$ as well. If $\langle v, v'\rangle_2 = 1$ then $2\langle v, v'\rangle_4 = 2$ and then $\langle w, w'\rangle_4 = 2$. This last case must occur. Otherwise $\mathcal{C}_X$ and $\mathcal{C}_Y$ would be self-orthogonal and, by Theorem 3, $\mathcal{C}$ would be separable. △

*Corollary 6:* Let $\mathcal{C}$ be a $\mathbb{Z}_2\mathbb{Z}_4$-additive self-dual code of type $(\alpha, \beta; \gamma, \delta; \kappa)$. If $\mathcal{C}$ is non-separable, then $\delta \geq 1$.

*Proof:* By Lemma 8, there are two codewords $(v, w), (v', w') \in \mathcal{C} \subseteq \mathbb{Z}_2^\alpha \times \mathbb{Z}_4^\beta$ such that $\langle v, v'\rangle_2 = 1$ and $\langle w, w'\rangle_4 = 2$. Therefore, either $w$ or $w'$ has a quaternary coordinate of order 4 and $\delta \geq 1$. ∎

*Theorem 7:* Let $\mathcal{C}$ be a $\mathbb{Z}_2\mathbb{Z}_4$-additive self-dual code of type $(\alpha, \beta; \gamma, \delta; \kappa)$, with $\alpha, \beta > 0$.

(i) If $\mathcal{C}$ is Type 0, then $\alpha \geq 2$, $\beta \geq 2$.

(ii) If $\mathcal{C}$ is Type I and separable, then $\alpha \geq 2$, $\beta \geq 1$.

(iii) If $\mathcal{C}$ is Type I and non-separable, then $\alpha \geq 4$, $\beta \geq 4$.

(iv) If $\mathcal{C}$ is Type II, then $\alpha \geq 8$, $\beta \geq 4$.

*Proof:* By Lemma 1, $\alpha = 2\kappa$. If $\mathcal{C}$ is Type 0, then there is a codeword $(x, y) \in \mathcal{C}$, where $wt_H(x)$ is odd and hence $p(y) \equiv 2 \pmod{4}$ by Lemma 3. Then, $\alpha \geq 2$ and $\beta \geq 2$.

If $\mathcal{C} = \mathcal{C}_X \times \mathcal{C}_Y \subseteq \mathbb{Z}_2^\alpha \times \mathbb{Z}_4^\beta$ is a Type I, separable code, then $\mathcal{C}_X$ is binary self-dual and $\mathcal{C}_Y$ is quaternary self-dual. Therefore, $\alpha \geq 2$ and $\beta \geq 1$.

Assume $\mathcal{C}$ is of Type I non-separable. Since $(\mathcal{C}_b)_X$ is self-dual and all codewords of $C_X$ are even weight, if $\alpha = 2$ then $(\mathcal{C}_b)_X = \mathcal{C}_X$, that is not possible. Hence $\alpha$ must be at least 4. Moreover, by Lemma 3 $p(y) \equiv 0 \pmod{4}$, for all $y \in \mathcal{C}_Y$. Since there is a codeword $y \in \mathcal{C}_Y$ with $p(y) \geq 1$ by Corollary 6, $p(y) \geq 4$ and $\beta$ must also be at least 4.

If $\mathcal{C} = \mathcal{C}_X \times \mathcal{C}_Y \subseteq \mathbb{Z}_2^\alpha \times \mathbb{Z}_4^\beta$ is a Type II, separable code, it is clear that $\mathcal{C}_X$ and $\mathcal{C}_Y$ have all weights doubly-even. Hence, $\alpha, \beta \geq 4$. If $\alpha = 4$, then $\mathcal{C}_X = \{0000, 1111\}$, but this is not a self-dual code. It is not possible $\alpha = 6$ because $\mathcal{C}_X$ should be antipodal and $\mathcal{C}_X$ would contain the vector $\mathbf{1}^6$ which is not doubly-even. Therefore, the minimum parameters are $\alpha = 8, \beta = 4$.

If $\mathcal{C}$ is Type II non-separable then we have that $\beta \geq 4$. By Corollary 5, we know that $\alpha + 2\beta$ is a multiple of 8. Therefore, for $\beta = 4$, we have $\alpha \geq 8$. ∎

Let $\alpha_{min}, \beta_{min}$ be the minimum values of $\alpha$ and $\beta$ given in Theorem 7 for each case $(i)$ to $(iv)$. Note that the codes $\mathcal{C}_1, \mathcal{C}_2, \mathcal{C}_3, \mathcal{C}_5$ and $\mathcal{C}_6$ are examples with the minimum values $\alpha_{min}$ and $\beta_{min}$.

*Theorem 8:* Let $\alpha_{min}$ and $\beta_{min}$ be as above.

(i) There exist a Type 0 or Type I code of type $(\alpha, \beta; \gamma, \delta; \kappa)$ if and only if $\alpha = \alpha_{min} + 2a$, $a \geq 0$, $\beta \geq \beta_{min}$.

(ii) There exist a Type II code of type $(\alpha, \beta; \gamma, \delta; \kappa)$ if and only if $\alpha = \alpha_{min} + 8a$, $\beta = \beta_{min} + 4b$, $a, b \geq 0$.

*Proof:* The necessary conditions are given by corollaries 1, 5 and Theorem 5. Moreover, we know that there exist a $\mathbb{Z}_2\mathbb{Z}_4$-additive code of type $(\alpha_{min}, \beta_{min}; \gamma, \delta; \kappa)$ for each case $(i)$ to $(iv)$.

Let $\mathcal{C}$ be a Type 0 or Type I code of type $(\alpha, \beta; \gamma, \delta; \kappa)$, with $\alpha = \alpha_{min} + 2a$, $a \geq 0$ and $\beta \geq \beta_{min}$. Consider the Type I codes $\mathcal{C}'$ and $\mathcal{C}''$ generated respectively by

$$\mathcal{G}' = \begin{pmatrix} 1 & 1 \mid \end{pmatrix}; \quad \mathcal{G}'' = \begin{pmatrix} \mid 2 \end{pmatrix}. \tag{6}$$

Then, by Lemma 6, $\mathcal{C} \times \mathcal{C}'$ is a $\mathbb{Z}_2\mathbb{Z}_4$-additive self-dual code of type $(\alpha + 2, \beta; \gamma + 1, \delta; \kappa + 1)$ and $\mathcal{C} \times \mathcal{C}''$ is a $\mathbb{Z}_2\mathbb{Z}_4$-additive self-dual code of type $(\alpha, \beta + 1; \gamma + 1, \delta; \kappa)$. If $\mathcal{C}$ is a Type 0 (resp. Type I) code then both codes $\mathcal{C} \times \mathcal{C}'$ and $\mathcal{C} \times \mathcal{C}''$ are Type 0 (resp. Type I).

If $\mathcal{C}$ is a Type II code of type $(\alpha, \beta; \gamma, \delta; \kappa)$, with $\alpha = \alpha_{min} + 8a$ and $\beta = \beta_{min} + 4b$, $a, b \geq 0$. Consider the Type II codes $\mathcal{C}'$, the extended Hamming code of length 8, and $\mathcal{C}''$ the code generated by

$$\mathcal{G}' = \begin{pmatrix} \mid 2 & 2 & 0 & 0 \\ \mid 2 & 0 & 2 & 0 \\ \mid 1 & 1 & 1 & 1 \end{pmatrix}. \tag{7}$$

By Lemma 6, $\mathcal{C} \times \mathcal{C}'$ is a Type II code of type $(\alpha + 8, \beta; \gamma + 8, \delta; \kappa + 8)$ and $\mathcal{C} \times \mathcal{C}''$ is a Type II code of type $(\alpha, \beta + 4; \gamma + 2, \delta + 1; \kappa)$. ∎

Finally, we remark a special case where the binary image is also a self-dual code.

*Theorem 9:* If $\mathcal{C}$ is a Type II code and $\Phi(\mathcal{C})$ is linear then $\Phi(\mathcal{C})$ is a binary doubly-even self-dual code.

*Proof:* If $v, w$ are binary vectors with $wt_H(v) \equiv wt_H(w) \equiv 0 \pmod 4$ then $wt_H(v+w) = wt_H(v) + wt_H(w) - 2\langle v, w \rangle_2$. Hence if $v + w \in \Phi(\mathcal{C})$ then it has doubly-even weight and hence $\langle v, w \rangle_2$ is even and hence they are orthogonal. Hence the code is a self-dual code. △

## V. WEIGHT ENUMERATORS OF $\mathbb{Z}_2\mathbb{Z}_4$-ADDITIVE SELF-DUAL CODES

Let $\mathcal{C}$ be a $\mathbb{Z}_2\mathbb{Z}_4$-additive self-dual code of type $(\alpha, \beta; \gamma, \delta; \kappa)$, $C = \phi(\mathcal{C})$. Since the weight enumerator of $C$ satisfies $W_{C^\perp}(x,y) = \frac{1}{|C|}W_C(x+y, x-y)$, then $C$ is held invariant by the action of the matrix:

$$M = \frac{1}{\sqrt{2}} \begin{pmatrix} 1 & 1 \\ 1 & -2 \end{pmatrix}. \tag{8}$$

The matrix $M$ satisfies $M^2 = I_2$.

We also know that the length of $C$, $n = \alpha + 2\beta$, is even so $W_C(-x,-y) = W_C(x,y)$ and so the weight enumerator is held invariant by the matrix:

$$B = \begin{pmatrix} -1 & 0 \\ 0 & -1 \end{pmatrix}. \tag{9}$$

*Theorem 10:* Let $\mathcal{C}$ be a $\mathbb{Z}_2\mathbb{Z}_4$-additive self-dual code. Then,

$$\begin{cases} W_\mathcal{C}(x,y) \in \mathbb{C}[x^2+y^2, y(x-y)], & \text{if } \mathcal{C} \text{ is Type 0,} \\ W_\mathcal{C}(x,y) \in \mathbb{C}[x^2+y^2, x^2y^2(x^2-y^2)^2], & \text{if } \mathcal{C} \text{ is Type I,} \\ W_\mathcal{C}(x,y) \in \mathbb{C}[x^8 + 14x^4y^4 + y^8, x^4y^4(x^4-y^4)^4], & \text{if } \mathcal{C} \text{ is Type II.} \end{cases} \tag{10}$$

*Proof:* If $G$ is the group $\langle M, B \rangle$ then $|G| = 4$ with $G = \{M, -M, B, I_2\}$. In [14], the invariant theory is done for this group. It has Molien series $\frac{1}{(1-\lambda^2)^2}$ and the weight enumerator of a $\mathbb{Z}_2\mathbb{Z}_4$-additive self-dual code belongs to $\mathbb{C}[x^2 + y^2, y(x - y)]$. For a Type I $\mathbb{Z}_2\mathbb{Z}_4$-additive code the usual Gleason's theorem for binary singly-even codes applies, namely a $\mathbb{Z}_2\mathbb{Z}_4$-additive Type I code has a weight enumerator in $\mathbb{C}[x^2 + y^2, x^2y^2(x^2 - y^2)^2]$. For a Type II $\mathbb{Z}_2\mathbb{Z}_4$-additive code the usual Gleason's theorem for binary doubly-even codes applies, namely a $\mathbb{Z}_2\mathbb{Z}_4$-additive Type II code has a weight enumerator in $\mathbb{C}[x^8 + 14x^4y^4 + y^8, x^4y^4(x^4 - y^4)^4]$. ∎

Consider the code given in Example 1. The vectors that have even weight are precisely

$$\{(00|00), (11|20), (11|02), (00|22)\}. \tag{11}$$

These form a linear subcode consisting of exactly half the vectors. We shall show that this holds in general.

*Lemma 9:* Let $\mathcal{C}$ be a Type 0 code then the subcode $\mathcal{C}_0 = \{\mathbf{v} \mid \mathbf{v} \in \mathcal{C}, wt(\mathbf{v}) \equiv 0 \pmod{2}\}$ is a linear subcode with $|\mathcal{C}| = 2|\mathcal{C}_0|$.

*Proof:* Let $v'$ and $w'$ be binary vectors then

$$wt_H(v' + w') = wt_H(v') + wt_H(w') - 2\langle v', w' \rangle_2, \tag{12}$$

Now let $v''$ and $w''$ be quaternary vectors. Recall that $\phi(v'' + w'') = \phi(v'') + \phi(w'') + \phi(2vw)$ where $vw$ is componentwise product. Then we have

$$\begin{aligned}
wt_L(v'' + w'') &= wt_H(\phi(v'' + w'')) = wt_H(\phi(v'') + \phi(w'') + \phi(2v''w'')) \\
&= wt_H(\phi(v'') + \phi(w'')) + wt_H(\phi(2v''w'')) \\
&\quad - 2\langle(\phi(v'') + \phi(w'')), \phi(2v''w'')\rangle_2 \\
&= wt_H(\phi(v'')) + wt_H(\phi(w'')) - 2\langle\phi(v''), \phi(w'')\rangle_2 \\
&\quad + wt_H(\phi(2v''w'')) - 2\langle(\phi(v'') + \phi(w'')), \phi(2v''w'')\rangle_2 \\
&\equiv wt_L(v'') + wt_L(w'') \pmod{2}.
\end{aligned}$$

Let $\mathbf{v}$ and $\mathbf{w}$ be vectors in $\mathcal{C}$ with even weight. Let $v', w'$ be the binary parts and $v'', w''$ be the quaternary parts of $\mathbf{v}$ and $\mathbf{w}$ respectively. Equation 12 and the previous gives that

$$wt(\mathbf{v} + \mathbf{w}) = wt_H(v' + w') + wt_L(v'' + w'') \equiv$$

$$wt_H(v') + wt_H(w') + wt_L(v'') + wt_L(w'') \pmod{2}.$$

Then the sum of $\mathbf{v}$ and $\mathbf{w}$ has even weight. Then there is a unique coset of $\mathcal{C}_0$ consisting of the odd weight vectors so $\mathcal{C}_0$ has index 2 in $\mathcal{C}$. $\triangle$

Note that for binary self-dual codes a similar thing is done except that $\mathcal{C}_0$ consists of doubly-even vectors. This notion cannot be extended here to Type I codes since the sum of two vectors with doubly-even weight may not have doubly-even weight.

*Lemma 10:* Let $\mathcal{C}$ be a Type 0 $\mathbb{Z}_2\mathbb{Z}_4$-additive code, then

$$W_{\mathcal{C}_0}(x, y) = \frac{1}{2}(W_\mathcal{C}(x, -y) + W_\mathcal{C}(x, y)).$$

*Proof:* Vectors have monomial representation $x^{n-wt(\mathbf{v})}y^{wt(\mathbf{v})}$, where $n = \alpha + 2\beta$. For odd weight vectors $x^{n-wt(\mathbf{v})}(-y)^{wt(\mathbf{v})} = -x^{n-wt(\mathbf{v})}y^{wt(\mathbf{v})}$. Hence the monomial in $W_\mathcal{C}(x, y))$ cancels with the monomial in $W_\mathcal{C}(x, -y))$. For even weight vectors $x^{n-wt(\mathbf{v})}(-y)^{wt(\mathbf{v})} = x^{n-wt(\mathbf{v})}y^{wt(\mathbf{v})}$, so the vectors are counted twice so dividing by 2 gives the monomial that represents each even vector. $\triangle$

We define the shadow of a $\mathbb{Z}_2\mathbb{Z}_4$-additive code $\mathcal{C}$ to be $S = \mathcal{C}_0^\perp$. The shadow is a non-linear code with $|S| = |\mathcal{C}|$. Recall that $M$ was the matrix that gave action of the MacWilliams relations.

Shadows of binary codes first appeared in [18] but were first specifically labeled as a code in [8]. The shadow has been generalized to numerous alphabets, see [17] for a complete description.

We can compute the weight enumerator of the shadow as follows:

$$\begin{aligned}
W_S(x,y) &= W_{\mathcal{C}_0^\perp(x,y)} - W_\mathcal{C}(x,y) \\
&= \frac{1}{|\mathcal{C}_0|} M \cdot W_{\mathcal{C}_0}(x,y) - W_\mathcal{C}(x,y) \\
&= \frac{1}{2|\mathcal{C}_0|} M \cdot (W_\mathcal{C}(x,y) + W_\mathcal{C}(x,-y)) - W_\mathcal{C}(x,y) \\
&= \frac{1}{|\mathcal{C}|} M \cdot W_\mathcal{C}(x,y) + \frac{1}{|\mathcal{C}|} M \cdot W_\mathcal{C}(x,y) - W_\mathcal{C}(x,y) \\
&= \frac{1}{|\mathcal{C}|} M \cdot W_\mathcal{C}(x,-y).
\end{aligned}$$

This gives the following.

*Theorem 11:* Let $\mathcal{C}$ be a Type 0 $\mathbb{Z}_2\mathbb{Z}_4$-additive code with shadow $S$ then

$$W_S(x,y) = W_\mathcal{C}\left(\frac{x+y}{\sqrt{2}}, \frac{-(x-y)}{\sqrt{2}}\right).$$

Notice the difference with the usual binary case in that these weight enumerators are not necessarily possible weight enumerators for binary self-dual codes since there can be odd weight vectors represented and that the shadow is computed differently. Given a possible weight enumerator for Type 0 codes one can compute the weight enumerator of the shadow.

*Example 6:* Returning to Example 1 we can compute the weight enumerators. Namely

$$\begin{aligned}
W_C(x,y) &= x^6 + 4x^3y^3 + 3x^2y^4 \\
W_{C_0}(x,y) &= x^6 + 3x^2y^4 \\
W_S(x,y) &= 3x^4y^2 + 4x^3y^3 + y^6.
\end{aligned}$$

The shadow $S$ consists of the vectors

$$\{(11|00), (01|11), (10|13), (00|20), (01|33), (00|02), (11|22), (10|31)\}. \tag{13}$$

All separable codes are Type I or Type II so a Type 0 code must be non-separable. It can however be decomposable since the direct product of two Type 0 codes is Type 0.

The code $\mathcal{C}_0$ has 4 cosets in $\mathcal{C}_0^\perp$. Let $\mathcal{C}_{0,0} = \mathcal{C}_0$ and $\mathcal{C}_{1,0} = \mathcal{C} - \mathcal{C}_{0,0}$. Let $\mathcal{C}_0^\perp = \mathcal{C} \cup \mathcal{C}_{0,1} \cup \mathcal{C}_{1,1}$, i.e. $S = \mathcal{C}_{0,1} \cup \mathcal{C}_{1,1}$. We denote the cosets like this since there must be vectors $\mathbf{s}$ and $\mathbf{t}$ with $\mathcal{C} = \langle \mathcal{C}_0, \mathbf{t} \rangle$ and $\mathcal{C}_0^\perp = \langle \mathcal{C}, \mathbf{s} \rangle$. Then we have that $\mathcal{C}_{i,j} = \mathcal{C}_0 + i\mathbf{t} + j\mathbf{s}$.

It can be chosen so that **s** has even weight then $\mathcal{C}_{0,0}$ and $\mathcal{C}_{0,1}$ consists of even weight vectors and $\mathcal{C}_{1,0}$ and $\mathcal{C}_{1,1}$ consists of odd weight vectors. Notice that $\mathbf{s} + \mathbf{s} \in \mathcal{C}$ so $\langle \mathbf{s} + \mathbf{s}, \mathbf{t} \rangle = 0$ which implies that $\langle \mathbf{s}, \mathbf{t} \rangle = 2$ since it cannot be 0. Of course, $\langle \mathbf{t}, \mathbf{t} \rangle = 0$ since it is a vector in a self-dual code. Note that $\mathbf{s} + \mathbf{s}$ is the sum of even vectors so by the proof of Lemma 9 it is an even vector. This implies that $\mathbf{s} + \mathbf{s} \in \mathcal{C}_0$. Moreover $\mathbf{t} + \mathbf{t} \in \mathcal{C}_0$ which gives that the group $\mathcal{C}_0^\perp / \mathcal{C}_0$ is isomorphic to the Klein-4 group.

Let **v** be a vector then $\langle \mathbf{v}, \mathbf{1}^\alpha \mathbf{2}^\beta \rangle = 2 wt(\mathbf{v}) \pmod 4$. This gives that the vector **v** is orthogonal to $\mathbf{1}^\alpha \mathbf{2}^\beta$ if and only if it has even weight. Therefore we can take $s$ to be $\mathbf{1}^\alpha \mathbf{2}^\beta$.

Then $\langle 1^\alpha 2^\beta, 1^\alpha 2^\beta \rangle = 2\alpha$. Since $\alpha$ is even we have that $\langle \mathbf{s}, \mathbf{s} \rangle = 0$. Then we have the orthogonality given in Table I.

TABLE I

ORTHOGONALITY RELATIONS

|  | $\mathcal{C}_{0,0}$ | $\mathcal{C}_{1,0}$ | $\mathcal{C}_{0,1}$ | $\mathcal{C}_{1,1}$ |
|---|---|---|---|---|
| $\mathcal{C}_{0,0}$ | 0 | 0 | 0 | 0 |
| $\mathcal{C}_{1,0}$ | 0 | 0 | 2 | 2 |
| $\mathcal{C}_{0,1}$ | 0 | 2 | 0 | 2 |
| $\mathcal{C}_{1,1}$ | 0 | 2 | 2 | 0 |

*Proposition 3:* Let $\mathcal{C}$ be a Type 0 code then the codes $\mathcal{C}_{0,0} \cup \mathcal{C}_{0,1} = \langle \mathcal{C}, \mathbf{s} \rangle$ and $\mathcal{C}_{0,0} \cup \mathcal{C}_{1,1} = \langle \mathcal{C}, \mathbf{s} + \mathbf{t} \rangle$ are self-dual neighbors of $\mathcal{C}$ that are not Type 0.

*Proof:* The result follows from the orthogonality table. △

We can now generalize the construction first described in [7] but greatly expanded in [11].

*Theorem 12:* Let $\mathcal{C}$ and $\mathcal{D}$ be Type 0 codes in $\mathbb{Z}_2^\alpha \times \mathbb{Z}_4^\beta$ and $\mathbb{Z}_2^{\alpha'} \times \mathbb{Z}_4^{\beta'}$ respectively. Let $\mathcal{C}_a$ and $\mathcal{D}_a$ be the $a$ given in Table I for each code respectively. If $\mathcal{C}_a = \mathcal{D}_a$ then the code

$$(\mathcal{C}_{0,0}, \mathcal{D}_{0,0}) \cup (\mathcal{C}_{0,1}, \mathcal{D}_{0,1}) \cup (\mathcal{C}_{1,0}, \mathcal{D}_{1,0}) \cup (\mathcal{C}_{1,1}, \mathcal{D}_{1,1})$$

is a self-dual code in $\mathbb{Z}_2^{\alpha+\alpha'} \times \mathbb{Z}_4^{\beta+\beta'}$. If $\mathcal{C}_a \neq \mathcal{D}_a$ then the code

$$(\mathcal{C}_{0,0}, \mathcal{D}_{0,0}) \cup (\mathcal{C}_{0,1}, \mathcal{D}_{1,1}) \cup (\mathcal{C}_{1,0}, \mathcal{D}_{1,0}) \cup (\mathcal{C}_{1,1}, \mathcal{D}_{0,1})$$

is a self-dual code in $\mathbb{Z}_2^{\alpha+\alpha'} \times \mathbb{Z}_4^{\beta+\beta'}$.

*Proof:* It is a simple computation to see that the vectors are orthogonal and that the code is linear. △

## VI. CONCLUSIONS

We summarize the main results of this paper in the following table.

|  | Type 0 | Type I | Type II |
|---|---|---|---|
| separable/ non-separable | non-separable | separable or non-separable | separable or non-separable |
| antipodality | non-antipodal | antipodal | antipodal |
| $W_\mathcal{C}(x,y)$ | $\mathbb{C}[x^2+y^2, y(x-y)]$ | $\mathbb{C}[x^2+y^2, x^2y^2(x^2-y^2)^2]$ | $\mathbb{C}[x^8+14x^4y^4+y^8, x^4y^4(x^4-y^4)^4]$ |
| separable $\alpha, \beta; a, b \geq 0$ | - - | $\alpha = 2 + 2a$ $\beta = 1 + b$ | $\alpha = 8 + 8a$ $\beta = 4 + 4b$ |
| non-separable $\alpha, \beta; a, b \geq 0$ | $\alpha = 2 + 2a$ $\beta = 2 + b$ | $\alpha = 4 + 2a$ $\beta = 4 + b$ | $\alpha = 8 + 8a$ $\beta = 4 + 4b$ |